\documentclass[epj]{webofc}
\usepackage[varg]{txfonts}   % Web of Conferences font
\wocname{EPJ Web of Conferences}
\woctitle{CONF12}
\begin{document}
\selectlanguage{english}
\title{Archeology and evolution of QCD}
%
% subtitle (optional, strongly discouraged)

\author{A. De R\'ujula\inst{1,2}\fnsep\thanks{\email{alvaro.derujula@cern.ch}} 
%\and
%        Second author\inst{2} \and
%        Third author\inst{3}
        % etc.
}

\institute{CERN
\and
           IFT/UAM, Madrid
%\and
%           The last address here
}

\abstract{%
These are excerpts from the closing talk at the
``XIIth Conference on Quark Confinement and the Hadron Spectrum'', 
which took place last Summer in Thessaloniki --an excellent place to enjoy 
an interest in archeology. 
A more complete personal view of the early days of QCD and the rest of 
the Standard Model is given in \cite{RefA}. 
Here I discuss a few of the points which
--to my judgement-- illustrate well the QCD evolution (in time), both from a scientific
and a sociological point of view.}
\maketitle
\section{Introduction}
\label{intro}
As one could judge from many talks at this conference \cite{Conf}, two areas of QCD have 
witnessed an enormous progress over the years. One of them is the non-perturbative
first-principle lattice calculation of many relevant observables.
The other embodies the next-to-next-to...leading-order results for many observed processes.
How do some of these results compare with the corresponding earliest ones?
How have the attitude and perceptions of ``the community'' evolved with time?

\section{Who invented quarks?}

The official history is that they were invented
by Gell-Mann and  Zweig, in that chronological order:
Gell-Mann's published paper was received by Physics Letters on January
4th 1964, while Zweig's unpublished work is a CERN yellow report dated
January 17th of the same year. But, as Napoleon is said to have said, 
{\it History is the version of past events that people have decided to agree
upon.}

According to the same un-trustable source of the previous quote 
(Internet) Gell-Mann's paper was originally rejected by Physical Review Letters.
Untrue \cite{MGM}.  None of this would significantly change the chronological order, 
which is anyway fairly irrelevant, the dates were so close. Concerning dates, it must be recalled 
that Gell-Mann wrote: {\it These ideas were developed ... in March 1963; the 
author would like to thank Professor Robert Serber for stimulating them.}
For Serber's recollections, see \cite{Serber}.

%The above history may be wrong in its dates, since Gell-Mann's paper
%was shamefully rejected by Physical Review Letters,  
%at least according to the same untrustable source of the previous quote 
%(Internet). That would not change the chronological order, which is anyway irrelevant;
% the dates were so close. But, concerning dates, it must be recalled 
%that Gell-Mann wrote: {\it These ideas were developed ... in March 1963; the 
%author would like to thank Professor Robert Serber for stimulating them.}

A point in the official history  \cite{Off} is lacking \cite{RefA,CC}.  Andr\'e Petermann
published a paper (in French!) \cite{Petermann},
received December 30th, 1963, shortly before the dates quoted
in the first paragraph.
In this paper he discusses mesons as made of a {\it spinor/anti-spinor
pair} and baryons as {\it composed of at least three spinors.} Concerning
the delicate issue of their charges, Petermann delightfully writes 
{\it if one wants to preserve charge
conservation, which is highly desirable, the spinors must have
fractional charges. This fact is unpleasant, but cannot, after all, be
excluded on physical grounds.}

There are other unofficial issues concerning this  chapter 
of the history of science. Was Zweig forbidden to have a preprint typed and to give a talk
at CERN at the time? If so, by whom? I shall not answer these questions,
but another one which I have been challenged to answer:
why was the publication of Petermann's paper delayed for a year?
Alas, nobody has found the original CERN preprint, yet.
So, one cannot disprove something evil I have been told, namely that {\it Petermann 
had plenty of time to change the paper before publication.}
That was not his style. Not only did he publish this paper in French
--guaranteeing a dearth of readers-- but in another uncontentious paper 
of his in the same journal, the editors clarify that the unusual
delay in the publication was not the fault of the journal.
The delay was seven years. The time needed by Petermann to bother correcting 
the proofs.

\section{Other quark mysteries, also unveiled}

The naive (constituent) quark model was impressively successful in its understanding of hadrons made of
$u$, $d$ and $s$ quarks and in predicting the existence, decay pattern and mass of the $\Omega^-$ \cite{M&Y}. But quarks and their confinement remained mysterious, more so because of the complementary evidence in SLAC's deep-inelastic electron-scattering experiments for charged constituents
of protons \cite{Dis}, Feynman's {\it partons} \cite{RF}, with ``point-like"  
interactions with photons --Bj{\o}rken's {\it scaling} \cite{BjF}.

I shall not discuss the original literature on Yang-Mills theories \cite{YM}, QCD \cite{GMFL},
the electro-weak standard model \cite{SLG,Steve,AS},
the necessary existence of charmed quarks \cite{GIM}
and the renormalizability of non-abelian gauge theories \cite{'t,'tV}. The discovery of 
strangeness-conserving neutral currents in neutrino scattering
by the Gargamelle bubble-chamber collaboration at CERN \cite{Musset} 
made experimentalists, and the world at large, aware of Yang-Mills theories, 
much as the 1971 work of 't Hooft \cite{'t} immediately attracted attention
from (field) theorists to the same subject. For the hypothetical young reader I must emphasize 
that the fact that the Standard Model had all the chances of being ``right"
was at the time only obvious to and up-to-then overwhelmed minority of field-theory addicts \cite{RefA}.

The understanding
of how quarks behaved when probed at short distances had to wait for the 
{\it discovery}\footnote{The remark, made also at this conference --that QCD's asymptotic freedom 
was first noticed by 't Hooft, Symanzik and the usual Russian suspects-- made me italicize ``discovery",
which includes the realization of how important something may be.}
 of QCD's asymptotic freedom \cite{DDF}. At the time David Politzer's office was next to mine at Harvard. 
 David Gross and Frank Wilczek were at Princeton. The Harvard/Princeton competition
 was acute \cite{DPNL} and productive \cite{RefA}.
 Suffice it to say that Harvard's
 motto is {\it VERITAS} (Truth), while Princeton's is {\it DEI SVB NVMINE VIGET} (God went to Princeton).
 
\section{A call for leniency }
\label{sec-1}

Hereafter I am going to cite papers in an unbalanced way, with a large fraction of references to papers authored or coauthored by me. Part of this is a proximity effect, I am writing as a witness and a $1/r^2$ law is inevitable, a price to pay for personal recollections, often more vivid than official histories.
%since {\it Memory... is the diary that we all carry with us}

As Golda Meyer put it: {\it Don't be humble... you are not that great}. That is correct in my case, but it does 
not apply to any of my to-be-cited coauthors, as the reader will easily recognize.

\section{ $\mathbf {\alpha_s}$ and $\mathbf{\Lambda}_{\rm\bf QCD}$}

In the mid 70's knives had been sharpened for long on inclusive reactions. 
The first concrete predictions of QCD  \cite{moments}
concerned the deviations from an exact scaling behaviour. But the available electron
scattering and $e^+e^-$ annihilation data \cite{Dis} covered
momentum transfers, $Q^2$, of not more than a few GeV$^2$. Nobody (yet) 
dared to analyse these data in the ``asymptotic'' spirit of QCD. But
some people not affected by dataphobia ---a morbid condition of the brain
(or brane?) that turns theoretical physicists into mathematicians--- set  out to
exploit the only data then available at higher $Q^2$.

By the early '70s, the proton's magnetic form factor $G_M$ had been measured
 \cite{proton} up to $Q^2\sim 20$ GeV$^2$. To bridge the gap between the QCD
predictions for deep inelastic scattering and an elastic form factor, two
groups  \cite{yo,GT} used (or, with the benefit of hindsight, slightly abused) the
then-mysterious ``Bloom--Gilman duality''  \cite{BG} relating the deep
``scaling'' data to the elastic and quasi-elastic peaks. 

Not atypical of the Harvard/Princeton
 competition of those times, the papers I just quoted
 \cite{yo,GT} were received by the publisher on
consecutive days. I prefer the one
beginning:  {\it ``Two virtues of asymptotically 
free gauge theories of the strong interactions are that they are {\bf not}
free-field theories and they make predictions that are {\bf not}
asymptotic''}; to conclude {\it ``The results agree with experiment but are
{\bf not} a conclusive test of asymptotic freedom.''} The first statement was --then-- not so obvious.

Being a bit more
inclined to data analysis than my Princeton competitors
--and wanting to be the first theoretical physicist [ (-: ]  to measure a fundamental constant of nature--
I extracted a value and an error range for 
$\Lambda_{\rm QCD}$,  while David Gross and Sam Treiman
simply chose a ``right'' value... somewhat surprisingly since
their results ---based on an analysis slightly different from mine---
neither fitted the data nor subtracted from their confidence in the theory
\cite{GT}. 

\begin{figure}[h]
\centering
\sidecaption
\includegraphics[width=10cm]{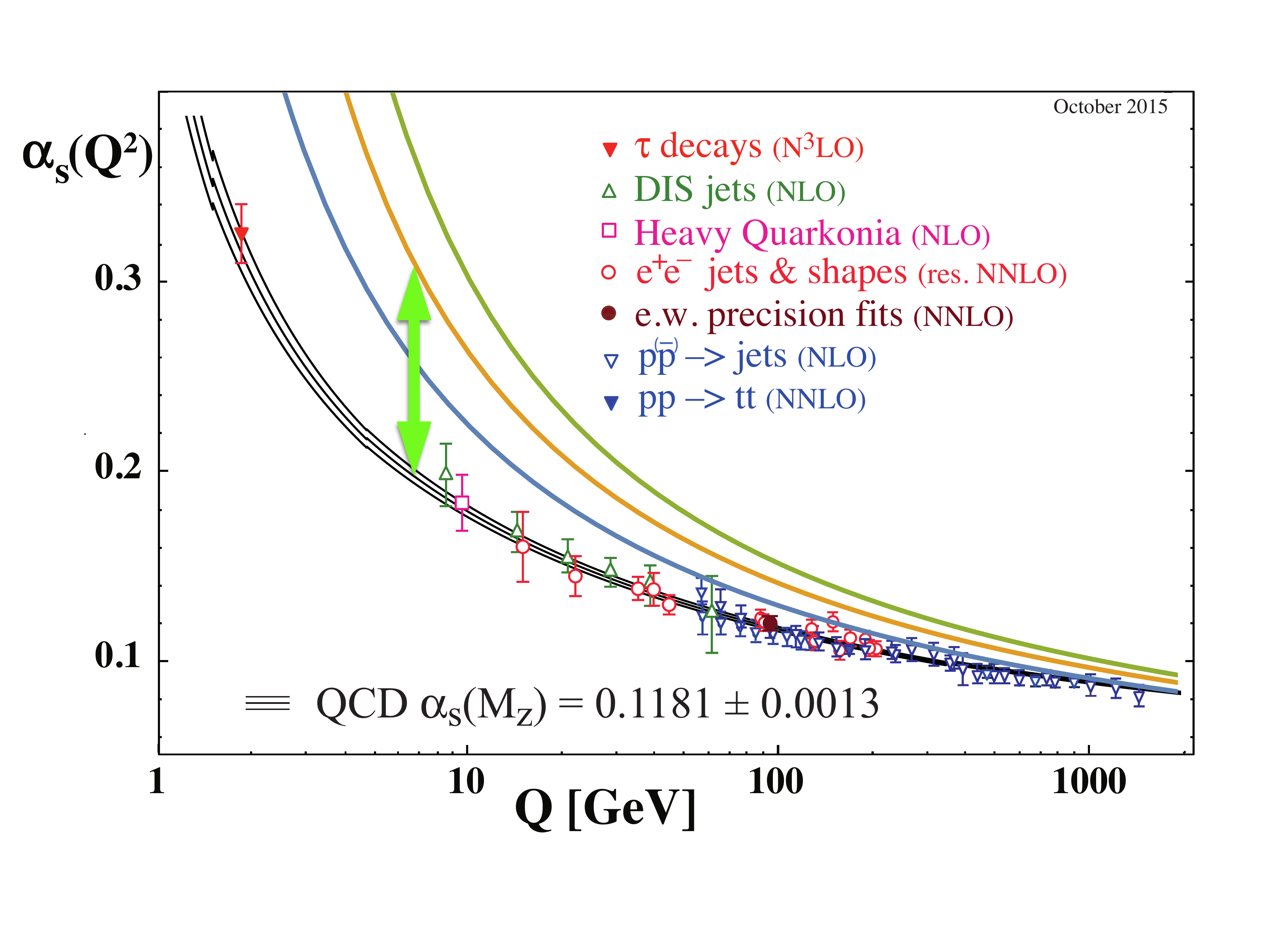}
\caption{Recent (black lines) and early (colored lines) results on $\alpha_s(Q^2)$. 
From \cite{alphas} and \cite{yo}, respectively.}
\label{fig-Lambda}     
\vspace{-1.5cm}
\end{figure}

In figure~\ref{fig-Lambda} I show recent and very sophisticated results \cite{alphas}
on QCD's fine-structure 
``constant", $\alpha_s(Q^2)$, as well as my original leading-order result, 
$\alpha_s=12/[25\,\pi\log(Q^2/ \Lambda^2)]$, and its error range. The vertical green double
arrow shows, tongue in cheek, that the central value of the original determination 
of $\alpha_s$ must be reduced by $\sim 33\%$ to get close to the current results.

\section{Bloom--Gilman duality}

I was asked at this conference whether Bloom--Gilman duality (BGD) is a prediction of QCD. It is not (yet).
That would require a complete understanding of bound-state production. But 
perturbative QCD {\it explains} BGD, in its QCD-improved realization. If you trust me, do not
read this technical section.

\begin{figure}[h]
%\hspace{-1cm}
\centering
\sidecaption
\includegraphics[width=12cm]{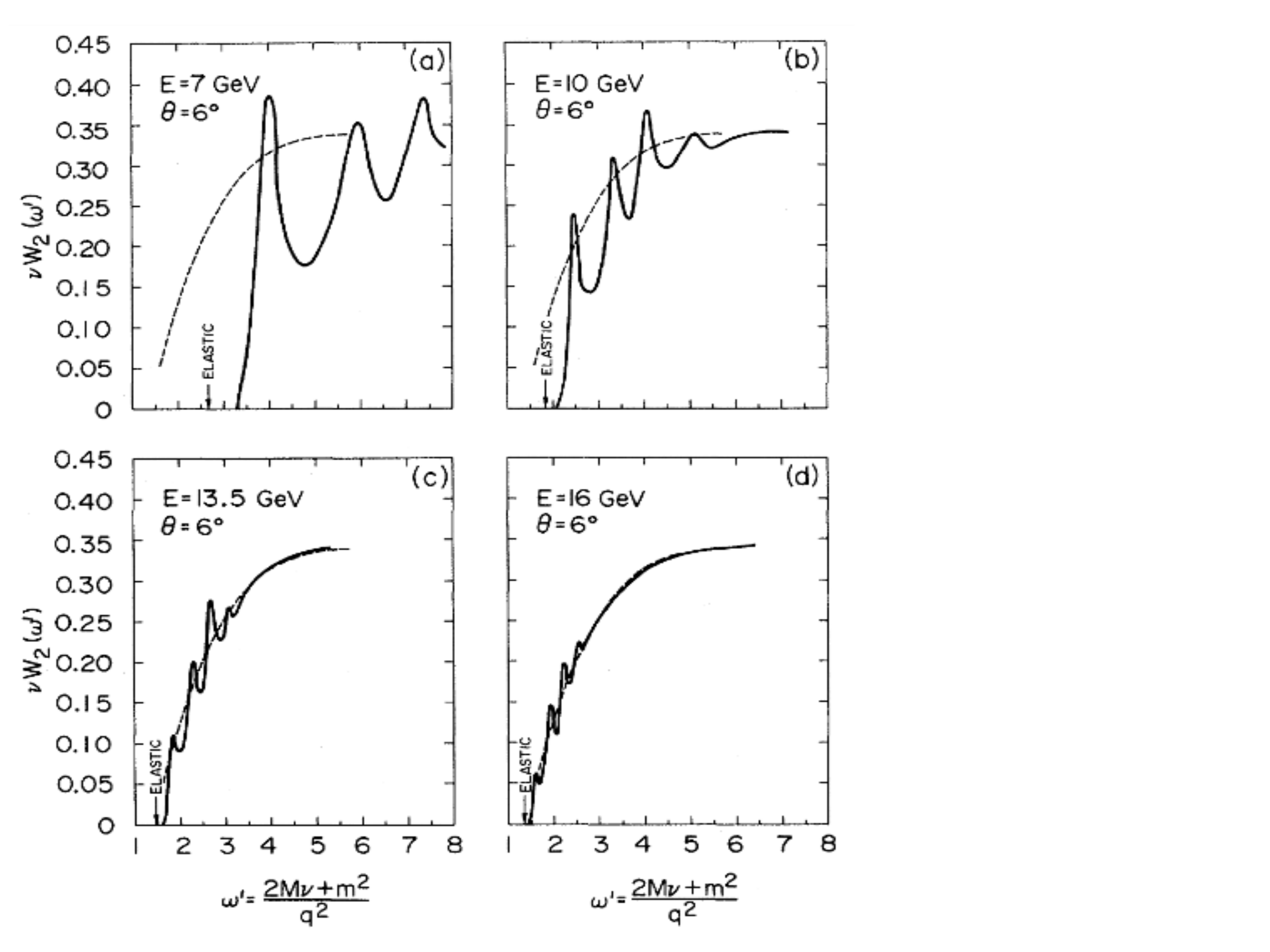}
\hspace{-3cm}
\caption{Bloom-Gilman duality at a fixed $e$-scattering angle and varying energies
(or $Q^2$ values).}
\label{fig-BG}       
\end{figure}

BGD is the observation \cite{BG} that at low $Q^2$ a structure function
shows prominent nucleon resonances, which ``average'' to the
``scaling'' function measured at some higher $Q_0^2$, and slide down
its slope as $Q^2$ increases. 
As shown in figure \ref{fig-BG}, this happens if the chosen scaling variable
in not Bj{\o}rken's $x=Q^2/(2\,m_p\,\nu)$, but contains a ``target mass correction",
$1/\omega' \equiv x'=Q^2/(2\,m_p\,\nu+m_p^2)$. All this was considered at the time,
in Californian style, as mystifying as a myth.

\begin{figure}[h]
\hspace{-1cm}
\centering
\sidecaption
\includegraphics[width=10.cm]{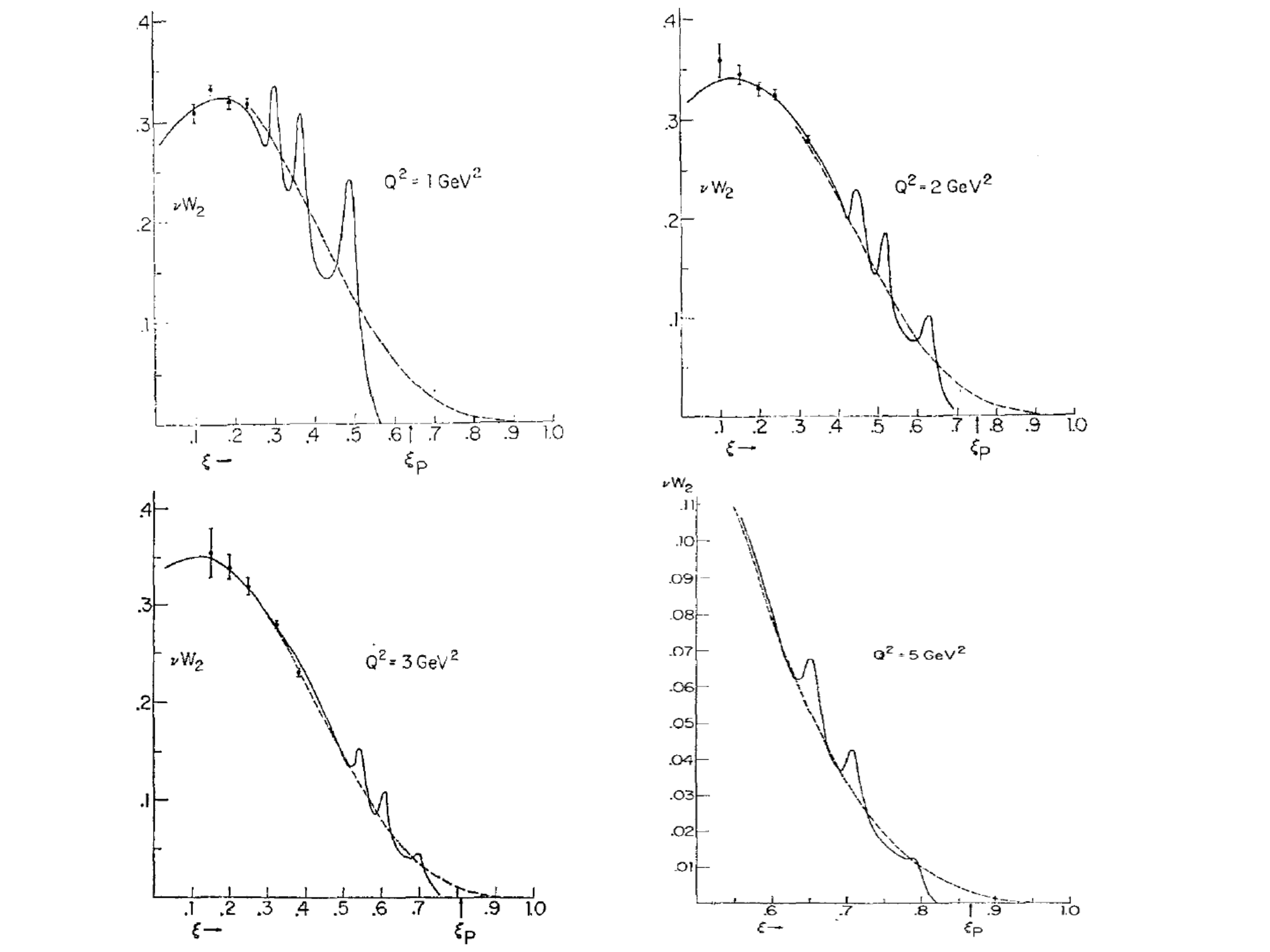}
\hspace{-1cm}
\caption{Dashed line: the perturbatively evolved proton structure function $\nu W_2(\xi,Q^2)$.
Continuous line: a fit to actual data. $\xi_p$ is the position of the elastic 
contribution $\propto G_{\!M}^2\delta(\xi-\xi_p)$. A few data points are also shown. }
\label{fig-BGD}       
\end{figure}

In  {\it Demythification of Electroproduction Local Duality and Precocious Scaling}
\cite{DGP2} Howard Georgi, Politzer and I argued that BGD is a consequence of QCD,
inevitable if scaling is ``precocious'', as it must be for small
$\Lambda$ (a fraction of a GeV). The scaling variable to be used, as we insisted in \cite{xi}, is
not $x'$, but the one implied by a full use of QCD's operator-product expansion, i.e.~Nachtman's 
variable $\xi= 2\,x/[1+(1+4\,m_p^2\,x^2/Q^2)^{1/2}]$ \cite{Otto}, which takes care of the target--mass
``higher-twist'' effects of leading order $m_p^2/Q^2$. In studying the $Q^2$-evolution of 
the $n$-th moment of  a structure function, weights $\xi^n$ and not $x^n$ ought to be used.
The entire structure function is to be $\xi^n$-weighed, including the elastic contribution at 
$\xi=\xi_p\equiv 2/[1+(1+4\,m_p^2/Q^2)^{1/2}]$. The QCD duality is shown in figure \ref{fig-BGD}.

The crucial point is that the customary logarithmic QCD evolution of structure functions
 has higher-twist corrections. The next to leading-twist ones are of the form 
$(1 + n\,a_n\, \Lambda^2/Q^2)$, with $|a_n|\simeq 1$, as the data allowed us to check in \cite{DGP2}.
Consider taking the $n$-th $\xi$-moment of a structure function measured at a 
relatively large $Q_0^2$, where the resonant peaks are barely observable.
Next, evolve this moment perturbatively down to a lower $Q^2\ll Q_0^2$.
Since the $a_n$ are not perturbatively calculable, the perturbative prediction for the moment
at the scale $Q^2$ has a relative uncertainty of ${\cal O}(n\,\Lambda^2/Q^2)$.

\begin{figure}[h]
\centering
\sidecaption
\includegraphics[width=7cm]{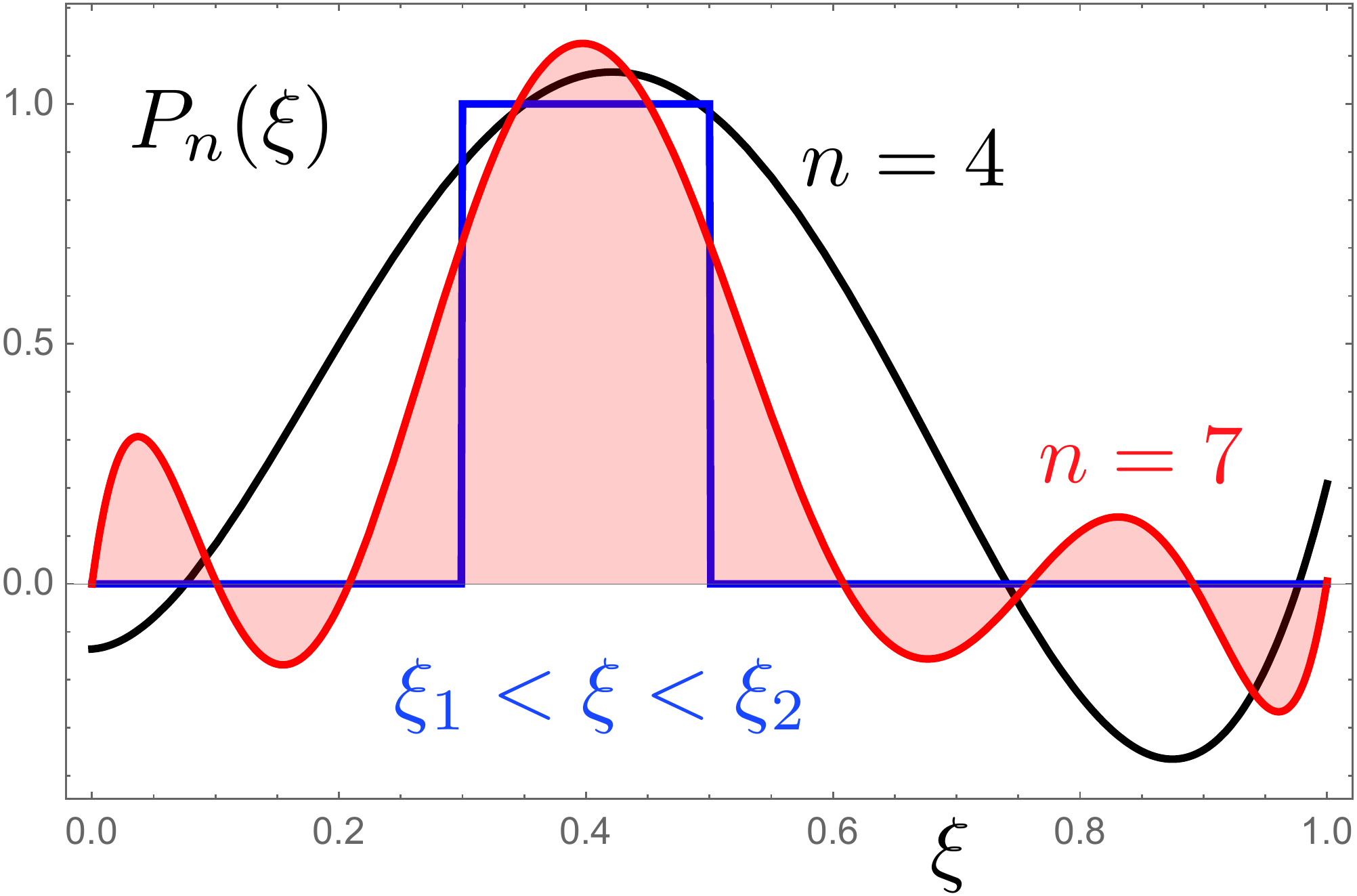}
\caption{Best fits to a window function from $\xi_1\,{\rm to}\; \xi_2$
with polynomials $P_n(\xi)$, for $n=4$ and 7.}
\label{fig-Polynomial}      
\end{figure}

To extract information on local duality from $n$ available moments, consider
 a polynomial $P_n(\xi)= \sum_0^n C_m\,\xi^m$.
One can find, for a given $n$, the $C_m$'s corresponding to a best fit to a ``window function"
that can be used to test duality ``locally'', i.e.~in a chosen interval $\xi_1\,{\rm to}\; \xi_2$, 
see figure \ref{fig-Polynomial}.
Given a predicted set of moments with uncertainties of ${\cal O}(n\,\Lambda^2/Q^2)$
one expects  a more local and precise QCD duality 
the smaller $n/Q^2$ is. That is precisely what is observed \cite{DGP2}. QED\footnote{
The actual analysis of QCD duality is a bit more elaborate, since one expects
slightly different precisions for window functions centered at different $\xi$'s \cite{DGP2}.}.

The preceding detailed discussion justifies a posteriori
the analysis of \cite{yo}, based on BG local duality in an interval
enclosing the elastic proton contribution $\propto G_{\!M}^2 \delta(\xi-\xi_p)$.
It also explains why those earliest attempts 
at QCD phenomenology resulted in a reasonable value for $\Lambda$.

\section{The irony of scaling in neutrino scattering}

The consensus that the observed scaling deviations smelled of QCD was
not triggered by theorists, but by an analysis of neutrino data by
Don Perkins {\it et al.}~\cite{Don}. This test of QCD was not very
severe. One reason is the neglect of higher twists  \cite{LM}.
Furthermore it is not possible, event by event, to measure the 
neutrino energy. Thus, in an unintended implementation of Bloom--Gilman
duality, a measured structure function, $F_\nu(x,Q^2)$, is significantly blurred in $x$ and $Q^2$.
This erases, at the
low $Q^2$ of a good fraction of the data, the 
resonance bumps that must be there, as in figure~\ref{fig-BGD}. Moreover, the total $\nu$
cross section increased linearly with $E_\nu$
 and was related to the naive, constituent-quark expectation from electroproduction by
a famous 18/5 factor, relating weak to electromagnetic quark charges.

Had the energy resolution of neutrino experiments
been as good as that of
their electron-scattering counterparts, the cross section rise would not have been so linear,
the nucleon-resonance bumps in the structure functions would have been
visible, and the data analysis would have had to be quite different.
With a pinch of poetic license one could assert that, early on, many
concluded that QCD was quite precise, but only  because the data were not.

\begin{figure}[h]
\centering
\includegraphics[width=13cm]{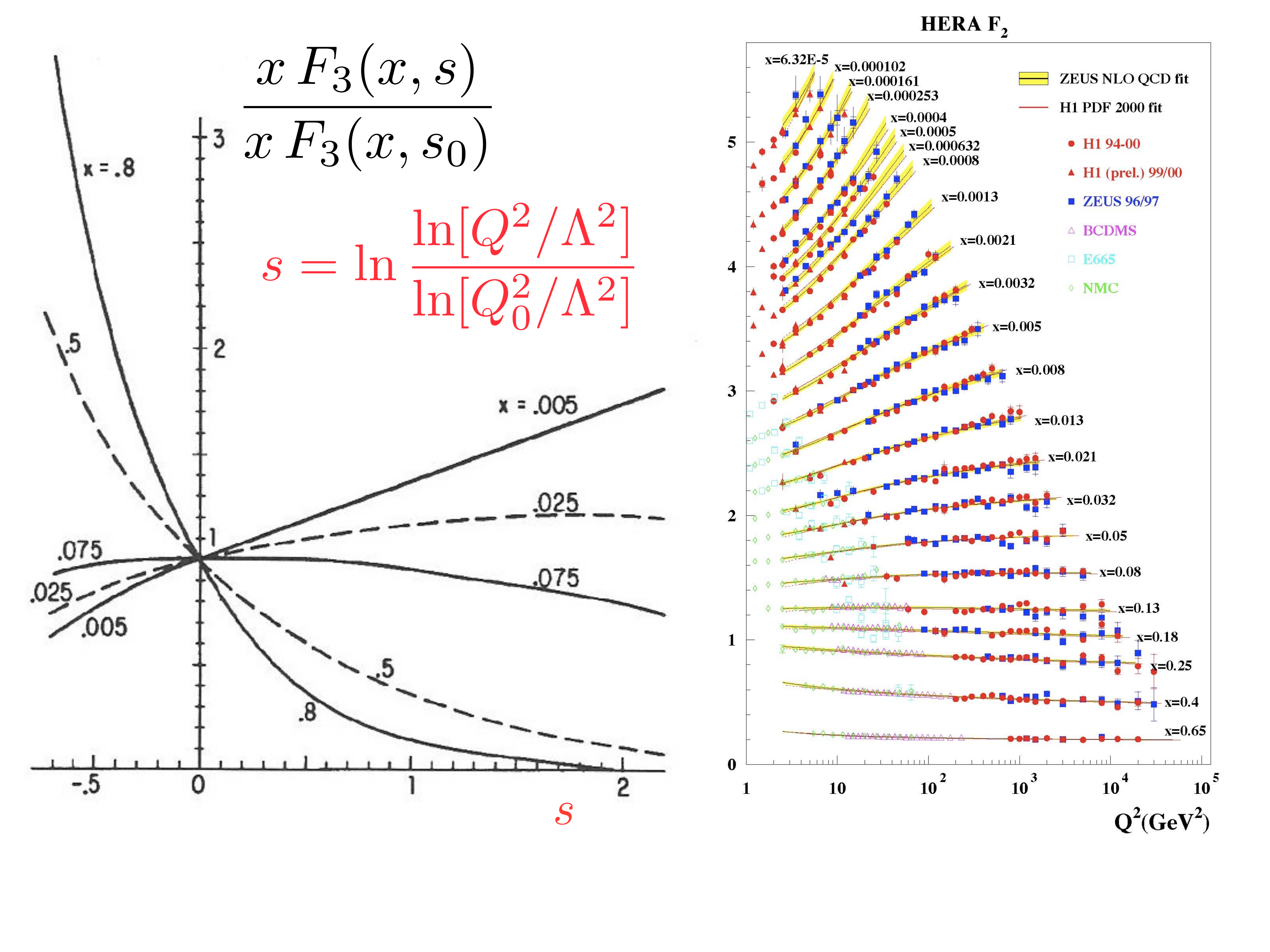}
\vspace{-1cm}
\caption{Left: Evolution of a normalized
 $\nu$ structure function $F_3(x,s)/F_3(x,0)$ at fixed 
$x$. The trend has been corroborated in detail by a
multitude of experiments (and theorists). Right: HERA electron-scattering data.}
\label{fig-SFS}      
\end{figure}

Elaborating on work by Giorgio Parisi \cite{Giorgio},
Georgi, Politzer and I explicitly worked out the $Q^2$ evolution of structure functions at 
fixed $x$, large $Q^2$ (for which $\xi\simeq x)$ \cite{DGP1}. The simplest results 
are for $x\,F_3(x,Q^2)$ (the C-odd neutrino scattering structure function), shown in 
figure \ref{fig-SFS}. A recent compilation of theory and HERA data for the electron scattering
$F_2(x,Q^2)$ is also shown in  the same figure.

The results shown in the left panel of figure~\ref{fig-SFS}
were to become heavily used...~and
systematically referenced to authors of later papers.
 While yowling, I plead guilty to having learned much later that
the simple underlying physics had been understood elsewhere: the
renormalization-group \cite{SP,GML}
picture of seeing partons within partons was drawn by
Kogut and Susskind  \cite{KS}, the ``physical gauge'' diagrammatic image of
a parton dissociating into others is due to Lev Lipatov
 \cite{Lip}, and its vintage QED analogue is none else than the
Weisz\"aker--Williams equivalent-photon approximation  \cite{WW}.

\section{The November Revolution}

\begin{figure}[h]
\centering
\sidecaption
\includegraphics[width=8cm]{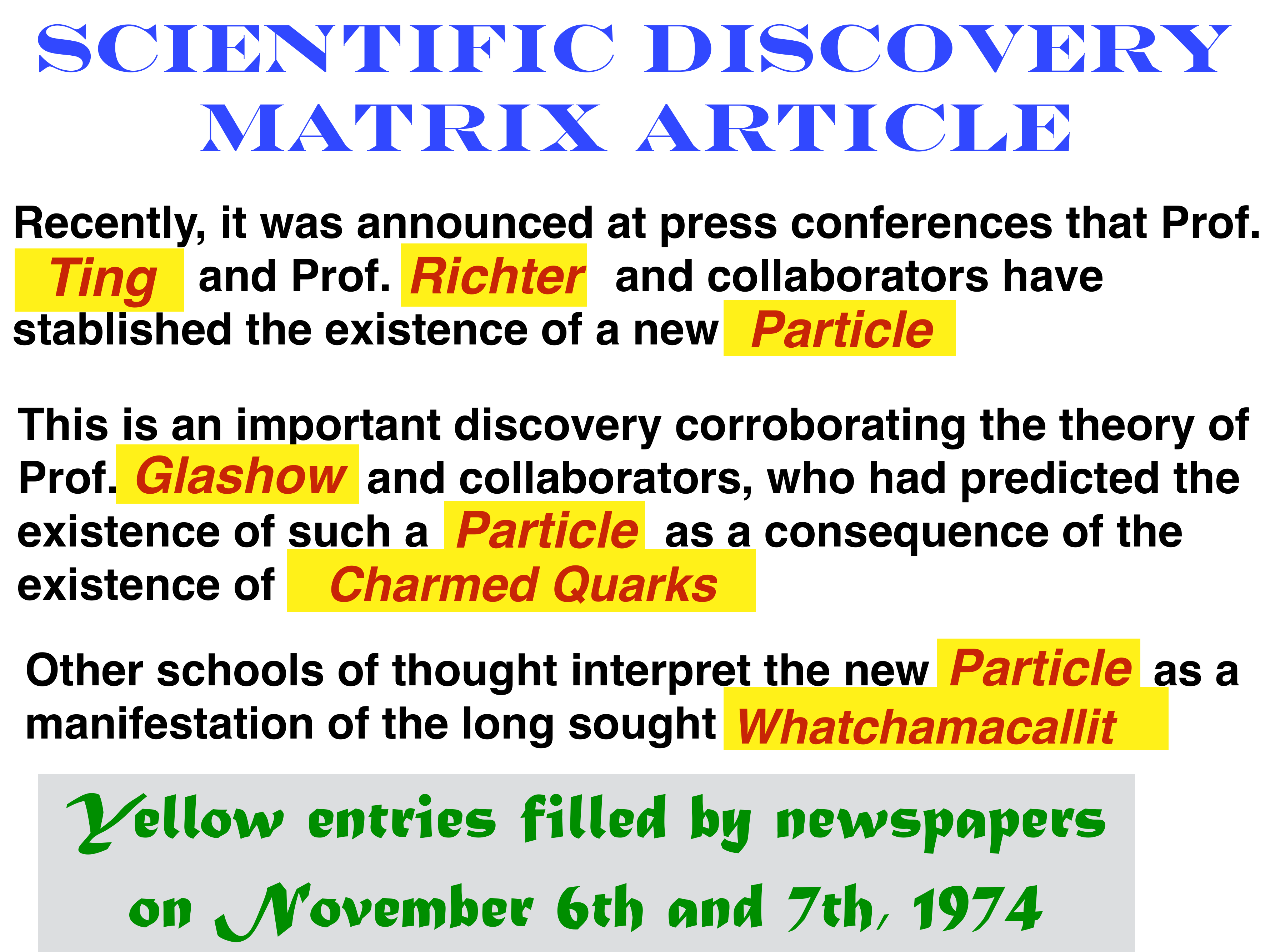}
\caption{The Scientific Discovery Matrix Article. The November 74 Revolution; fully filled version.}
\label{fig-SciDis}      
\end{figure}

In the 1970's the pace of discovery was so fast that (lazy) journalists decided to prepare
a ``matrix article'', wherein only the details of each specific discovery had to be filled in
at the last minute. The NYT matrix article and its filling in November 1974 are shown
in figure~\ref{fig-SciDis}.

Ten days of November 1974 shook the world of physics. 
Something wonderful and {\it almost}  \cite{AP} unexpected was to see the
light of day:  a  very discreetly charmed particle
 \cite{J,Psi}, a hadron so novel that it hardly looked like one.
Thirty years later, it is not easy to  recall  the 
collective ``high'' in which this discovery, and others to be made
in the two consecutive years, submerged the particle-physics community.
In my opinion, a detailed account that reflects well
the mood of the period is that by Riordan  \cite{Riordan}. In a
nutshell, the standard model arose from the ashes of the November
Revolution, while its competitors died honourably on the battleground.

\begin{figure}[h]
\centering
\sidecaption
\includegraphics[width=10cm]{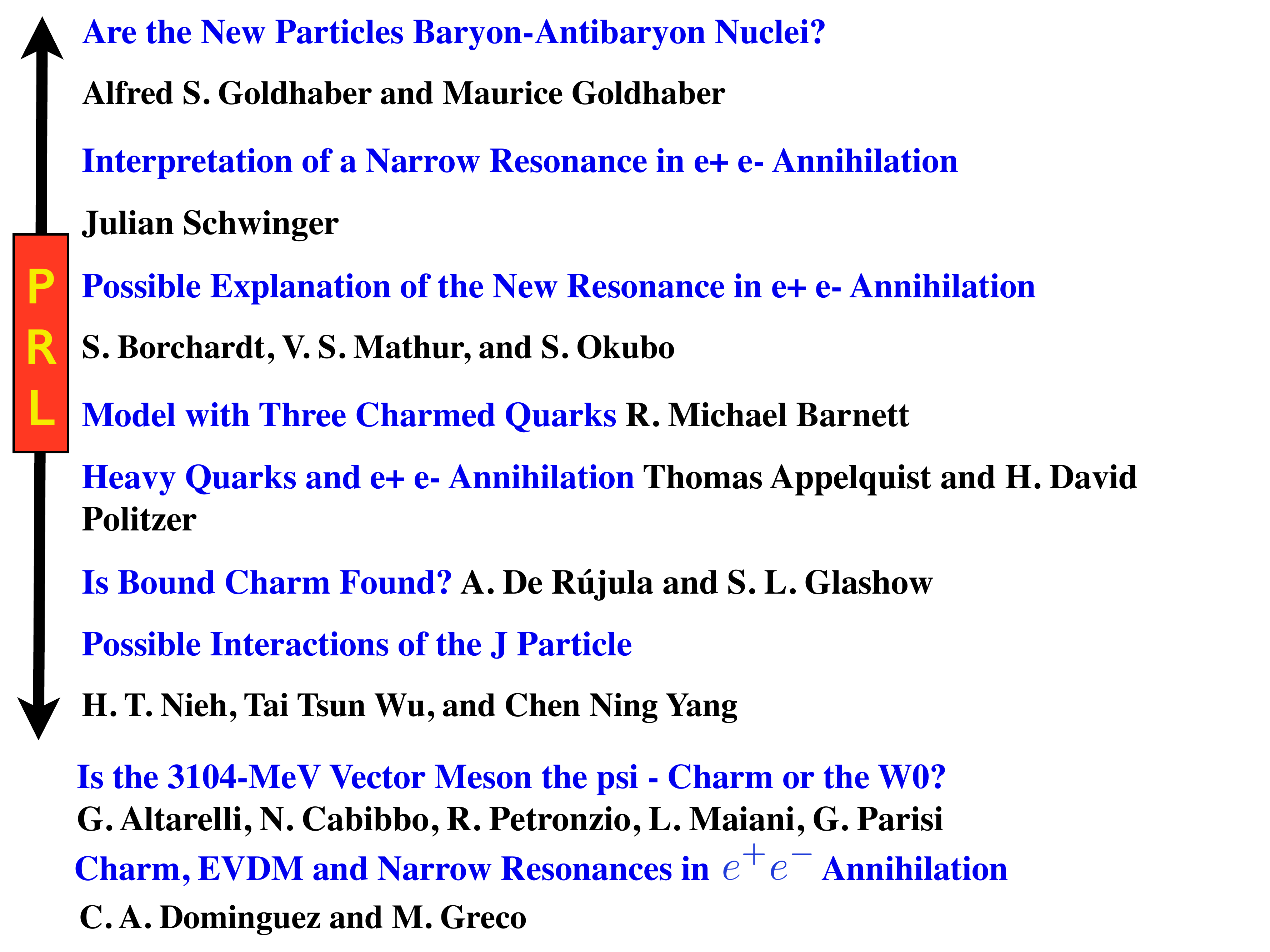}
\caption{Immediate interpretations of the $J/\psi$, with their titles.
PRL is Phys. Rev. Lett. {\bf 34}, Jan.~6th, 1975.
The last two papers are in Lett. Nuovo Cim. \cite{Romans,Mario}}
%{\bf 11},  609 and {\bf 12}, 439 (1975),
% respectively.}
\label{fig-Interpretations}      
\end{figure}

A couple of survivors and many of the casualties are shown in figure~\ref{fig-Interpretations}.
All of them but the last two were published, {\it unrefereed,} in the same issue of PRL.
%Volume 34, January 6th, 1975. 
The comparison of the $J/\psi$ and its interpretations 
to the  ones concerning the
recent (un)discovery of the
750 GeV state is... finding the proper adjective is left as an exercise for the reader,
or for the hoards of referees.

In the early Fall of '74, Tom Appelquist and David Politzer 
had been looking leisurely at how asymptotic freedom could imply a
positronium-like structure for the $c\bar c$ bound states of a
charmed quark and its  anti-goodie-bean. In those days, both QCD and
charm were already fully ``established'' at Harvard.
%, where one could,
%without the scorn of one's peers, violate the first rule of scientific fairy tales:
%not more than one fancy character per tale! 
Since Americans are often
short of vocabulary, my first contribution to the
subject was to baptize their toy {\it charmonium}. David and Tom's
first charmonium spectrum was so full of  Coulomb-like peaks  that they
could not believe it themselves. They debated the problem long enough
for the experimental avalanche to catch up with them. It was  a heavy 
price to pay for probity.

%Burt Richter was also caught in the avalanche. 
%On a short visit to Harvard, and with a healthy disrespect for theory, Burt told
%us that the electron spent some of its time as a hadron. 
%In answer to a question by Tom, he explained that sufficiently narrow
%resonances would escape detection in $e^+e^-$ colliders. Nobody around was aware of the possibility of catching the
%devil by its radiative tail (the emission of photons by the colliding
%particles widens the observed resonance on its $\sqrt{s}>M$ side). 
%Our vain discussions came to an abrupt end; 
%a rather urgent call summoning Burt back to SLAC delivered us from his 
%scorn for theorists. Only in other
%multi-world histories of our Quantum Universe  \cite{otroyo} do
%charmed theorists get to talk also with Sam Ting, prior to the
%Revolution\footnote{Sam did talk to a theorist at MIT, but Harvard's
%charmed infatuation did not extend that far.}.

For an object of its mass, the $J/\psi$ is four orders of magnitude
narrower than a conventional hadron resonance, and one order of magnitude
wider than a then hypothetical weak intermediary. It could not be either.
% (only because they are my good old
%friends do some people escape reference and derision at this point).
 Of the multitude of theoretical papers of figure~\ref{fig-Interpretations},
% \cite{ava}, 
 only two attributed the narrow width to asymptotic
freedom, one by Tom and David  \cite{AP}, who had intuited the whole
thing before, the other one by Sheldon Glashow and me  \cite{DeRG1}. I
recall Shelly storming the Lyman/Jefferson lab corridors with the notion of the
feeble three-gluon hadronic decay of the $J^P=1^-$ {\it
orthocharmonium} state, and I remember Tom and David muttering:
``Yeah''. Our paper still made it to the publishers in the auspicious
November, but only on the 27$^{\rm th}$, a whole week after the
article of our Harvard friends. The paper by Cesareo Dominguez and Mario Greco
also singled out charmonium as the interpretation.

We did a lot in our extra week  \cite{DeRG1}. {\it Abusus non tollit
usum} (of asymptotic freedom) we related the hadronic width of the
$J/\psi$ to that of $\phi \to 3\,\pi$, to explain why this hadronic resonance
 {\bf had to be} so narrow. We correctly estimated the
yields of production of truly charmed particles in $e^+e^-$
annihilation, $\nu$-induced reactions, hadron collisions and
photoproduction. Our mass for the $D^*$ turned out to be $2.5 \% $
off, sorry about that. We even discussed
mass splittings within multiplets of the
same quark constituency as {\it hyperfine}, a fertile notion. In
discussing paracharmonium ($J^P=0^-$) we asserted that {\it ``the
search for monochromatic $\gamma$'s should prove rewarding''}.
Finally, we predicted the existence of $\psi '$, but this time it was
our turn to be overtaken by the pace of discovery. 
%Not every week of
%my (scientific) life parallels this particular one. 
%The commentary on
%this paper in the book by Riordan \cite{Riordan} is one of those things
%that my grandfathers would have liked to read. My grandmothers may
%even have believed it!

\section{Charmonium spectroscopy}

I have a few vivid printable recollections of the times I am
discussing. One  concerns the late night in which the existence of
$P$-wave charmonia hit my head: we had been talking about $L=0$
states without realizing (we idiots!) that a bunch of $L=1$ charmonia
should lie between $J$ and $\psi '$ in mass. Too late to call Shelly,
 I spent hours guessing  masses and estimating the obviously
all-important $\gamma$-ray transition rates. At a gentlemanly morning
hour I rushed on my bicycle to Shelly's office, literally all the way
in, and attempted to snow him with my findings. I was speachless, out of breath
and wits. Shelly profited
to say: ``I know exactly what you are trying to tell me, there are
all these $P$-wave states etc., etc.'' He had figured it all out at
breakfast. I hated the guy's guts.

\begin{figure}[h]
\centering
\includegraphics[width=14cm]{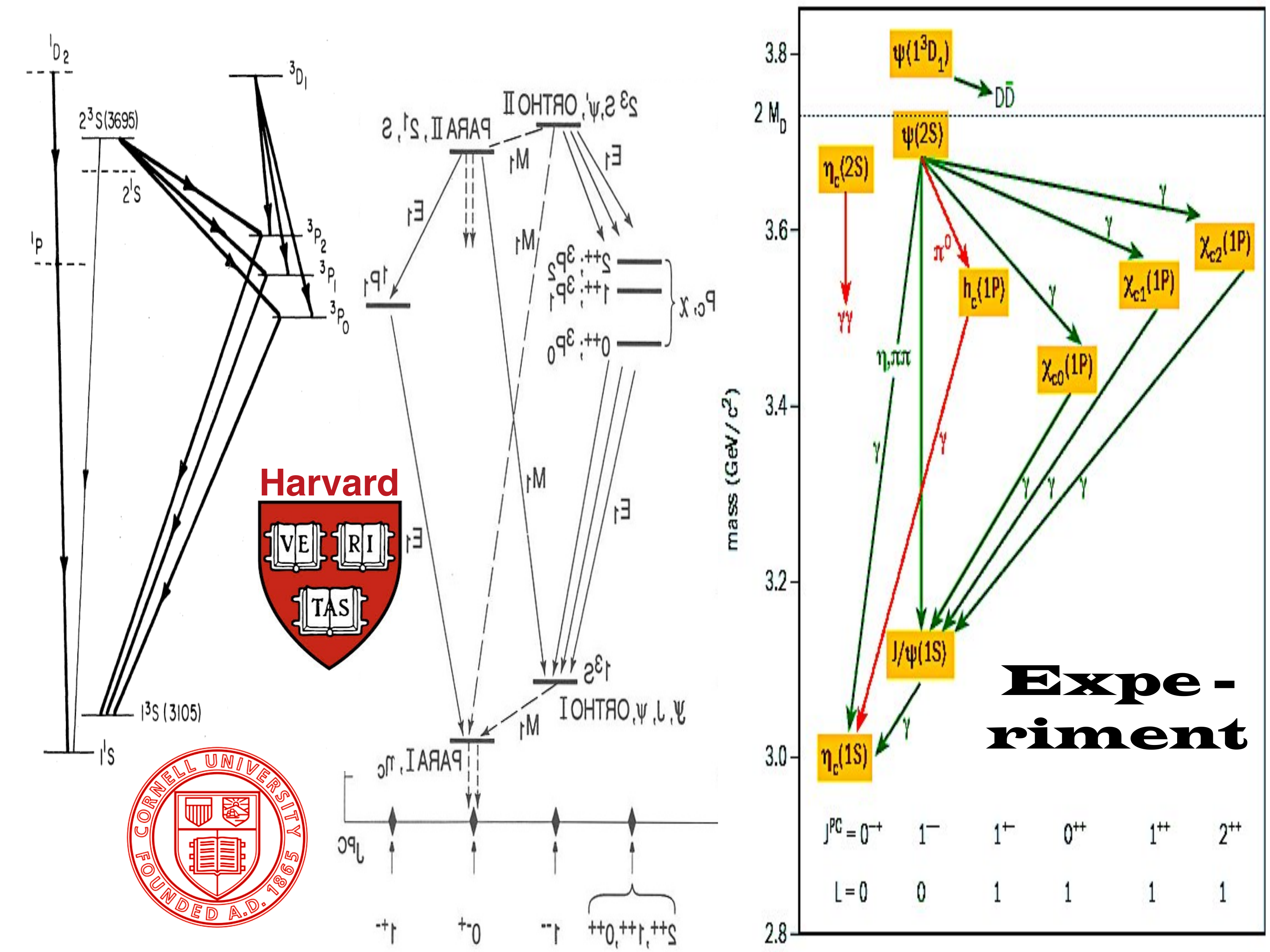}
\caption{The spectra of charmonia. From left to right: Cornell's (squeezed), Harvard's (mirror reflected) 
and  observed (with the inclusion of some non-radiative decays).}
\label{fig-Charmonia}      
\end{figure}

In no time, David and Tom gathered forces with Shelly and me to produce
an article  \cite{ADGP} on {\it Charmonium spectroscopy.} Physical
Review Letters  was fighting its usual losing battle against
progress (in nomenclature, $\smile$) and did not accept the title. 
Neither did PRL accept a similar title by our Cornell competitors \cite{Cornell}.
The predicted spectra and the current experimental situation are shown
in figure~\ref{fig-Charmonia}.

We, {\it the Crimson}, estimated the energy levels as ``half-way" between those
of a Coulomb and a harmonic oscillator potential. Indeed, a linear potential
--adequate for confined $c\bar c$ states-- is somewhat half-way. Our Cornell
friends, {\it the Carnelian}, borrowed a linear-potential program from Ken Wilson
and got similar results. Except for the all-important $\gamma$-ray transition
rates, for which the Carnelian predictions were much better than ours.

\section{Hadron masses in a gauge theory}

Early in 1975, Georgi, Glashow, and I wrote a paper \cite{DGG} whose style
reflects how high we rode, as well as how unorthodox QCD still was.
But for the added parentheses, here is how it began:

{\it Once upon a time, there was a controversy in particle physics.
There were some physicists who denied the existence of structures
more elementary than hadrons, and searched for a self-consistent
interpretation wherein all hadron states, stable or resonant, were
equally elementary} (the bootstrap). {\it Others, appalled by the teeming democracy of
hadrons, insisted on the existence of a small number of fundamental
constituents} (quarks) {\it and a simple underlying force law} (QCD). {\it In terms of these
more fundamental things, hadron spectroscopy should be qualitatively
described and essentially understood just as are atomic and nuclear physics.}

To the non-relativistic  quark model, we
added chromodynamic interactions
entirely analogous to their electrodynamic counterparts. We shall see that to this day
it is not totally clear why the ensuing predictions were so good. Our paradigmatic
result was the explanation of the origin and magnitude of the
$\Sigma^0$--$\Lambda$ mass difference. The two particles have the same
spin and quark constituency, their mass difference is a {\it
hyperfine} splitting induced by spin--spin interactions between
the constituent quarks. A little later, the ``MIT bag'' community published their 
relativistic version  \cite{MIT} of the same work.

In  \cite{DGG} we also predicted the masses of all ground-state
charmed mesons and baryons and (me too, I'm getting bored with this)
we got them right on the mark. Predictions based on an
incredible SU(4) version of the Gell-Mann--Okubo SU(3) mass formula,
and also the more sensible bag results, turned out to be wrong.
Only one person --indeed, again a Russian-- trusted a ``QCD-improved'' 
constituent quark model early enough to make predictions somewhat
akin to ours: Andrei Sakharov  \cite{Sach}.

\subsection{Good News at Last}

While theorists faithfully ground out the phenomenology of QCD,
experimentalists persistently failed to find decisive signatures of our
Trojan horse: the charmed quark. At one point,  the upper limits on the
$\gamma$-ray transitions of charmonia were well below the theoretical
expectations. Half of the $e^+e^-$ cross-section above $\sqrt{s}=4$ GeV was
due to charm production, said we. Who would believe that experimentalists
couldn't tell?

In the winter of '75 we saw a lone ray of light. As Nick Samios
recalls in detail in \cite{NSamios}, 
a Brookhaven bubble-chamber group  \cite{Nick} pictured
a $\Delta S=-\Delta Q$ event, forbidden in a charmless world, and
compatible with the chain:
\begin{equation}
\nu_\mu\;p  \rightarrow \Sigma^{++}_c \mu^-\, ; \;\;\;
\Sigma^{++}_c  \rightarrow \Lambda^+_c\; \pi^+\, ;  \;\;\;
\Lambda^+_c  \rightarrow \Lambda\;\pi^+\;\pi^+\;\pi^- \; .\nonumber
\end{equation}
Two charmed baryons discovered
in one shot!  This was a
source of delight not only for us, but also for the experimentalists
involved. They deserve a reproduction of their event: figure~\ref{fig-NS}, and a quotation:

\begin{figure}[h]
\centering
\sidecaption
\includegraphics[width=8cm]{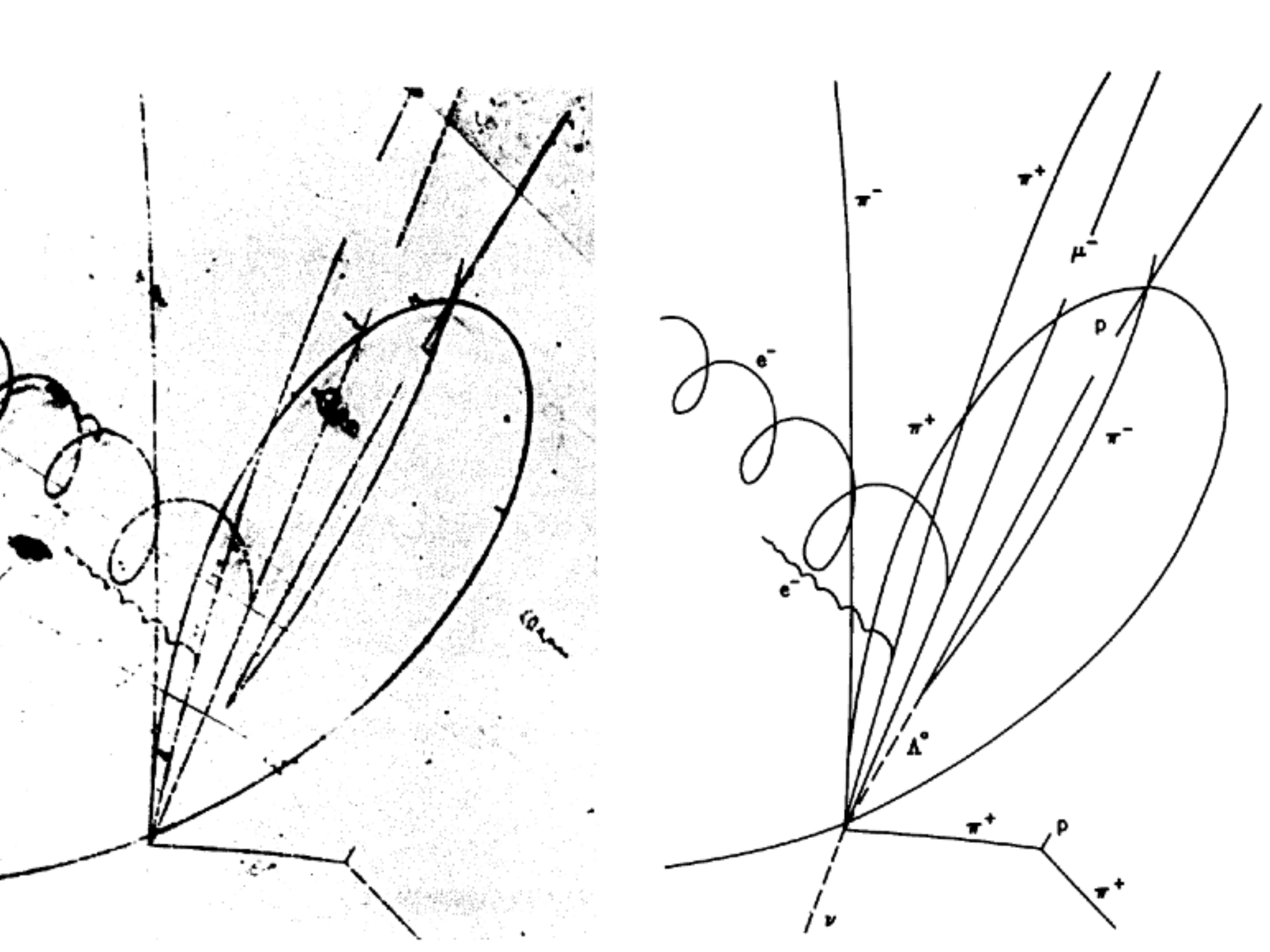}
\caption{The Brookhaven doubly charming event \cite{Nick}.}
\label{fig-NS}       
\end{figure}

{\it The total recoiling hadron mass ($\Lambda\,\pi^+\pi^+\pi^-$)} [is] $2426\pm 12$ {\it MeV.}

{\it This mass is in reasonable} {\it agreement with the value predicted by De R\'ujula, Georgi
and Glashow for the lowest-lying charmed-baryon states of charge +2, $2420$ MeV}
($J^P={3 \over 2}^+,\; I=1,\;\Sigma_c^*$) ... {\it There are three $\pi^+$'s and thus three mass 
differences derivable form this event; these
are observed to be $166\pm15$ MeV, $338\pm12$ MeV, and $327\pm 12$ MeV. The first of
these differences is in remarkable agreement with the $160$ MeV predicted for the decay of
a spin-$1\over 2$ charmed baryon $\Sigma_c$ decaying into a charmed $\Lambda_c$.}

This is {\it almost} precisely the way I feel experimentalists should write papers. 
Only ``almost'' because
the agreement between $2426\pm 12$ and 2420 MeV seems to me to be a bit better
 than ``reasonable''.

\subsection{Back to the future}  

\begin{figure}[h]
\centering
\sidecaption
\includegraphics[width=10cm]{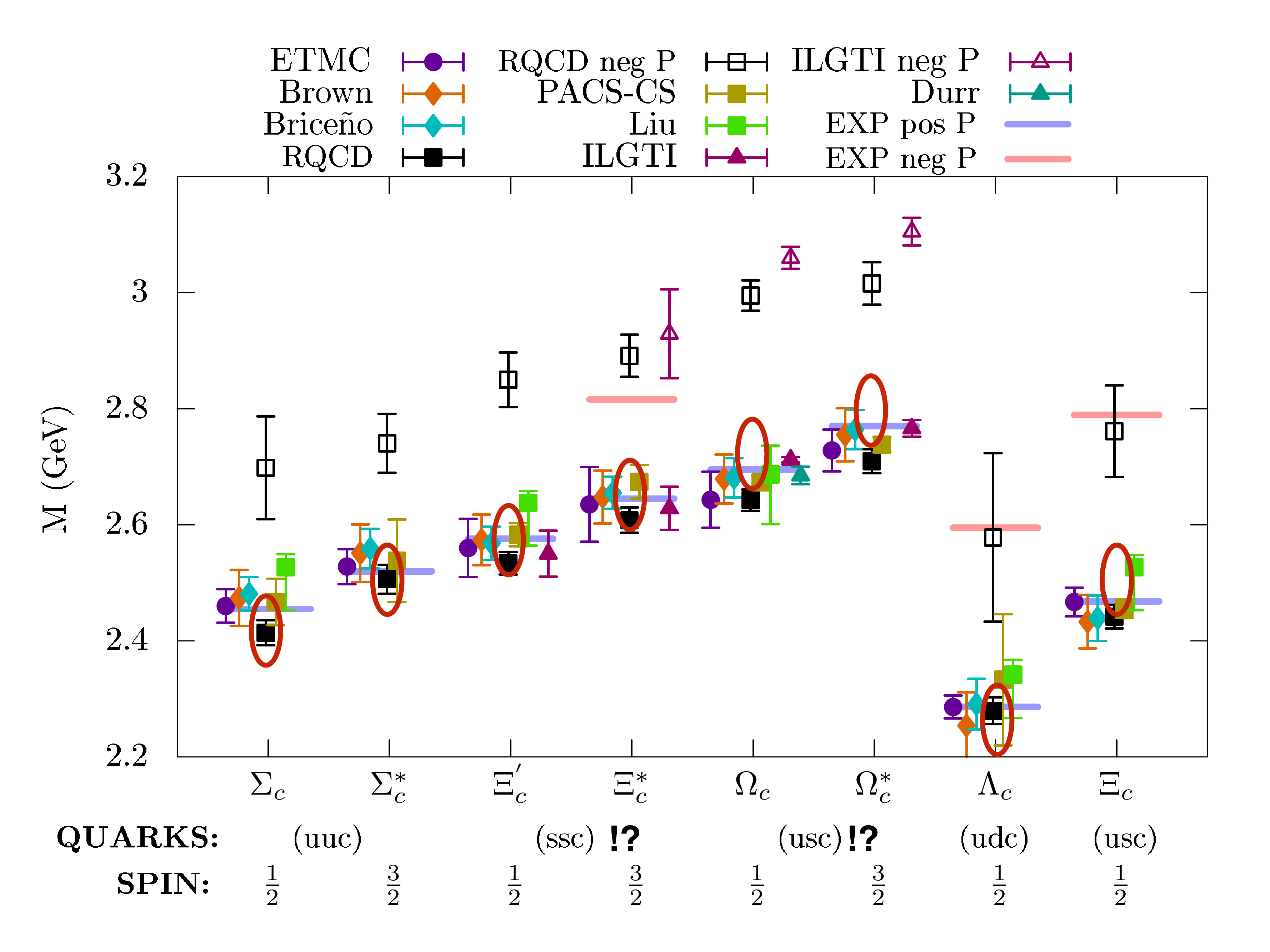}
\caption{Charmed baryons. Blue traits: mass measurements for the positive parity
ones.  Lattice results compiled in \cite{PCB} [(ssc) and (usc)
are interchanged]. Ellipses: the predictions of \cite{DGG},
with their common uncertainty.}
\label{fig-CBs}      
\end{figure}

By now the masses of many mesons and baryons have been precisely post-dicted
in lattice QCD. It is perhaps instructive to look at an instance:  charmed baryons.
This is done in figure~\ref{fig-CBs}, where a collection of lattice results \cite{PCB}    
is shown, along with the observed values (blue lines) for the positive parity baryons
we are concerned with. Also shown in the figure are the predictions of \cite{DGG},
made after the discovery of the $J/\psi$, but before that of open charm.
This limited information made us (over)estimate a common uncertainty of
$\pm 50$ MeV --reflected as the ellipses-- around the central values.

The QCD-improved naive quark model predictions of \cite{DGG} still compete
with the lattice results. The moral is that there is an element of ``truth'' in the
naive model, or in quenched lattice results. 
Presumably, if latticists could device a gauge-invariant way to characterize
a confined constituent-quark propagator, it would be found to peak at a
``constituent mass'' some $300$ MeV larger than a ``Lagrangian'' or chiral-model 
mass\footnote{This happens in Bethe-Saltpeter models
of confinement \cite{CG}. I am indebted to Pilar Hernandez for input on
this topic.}.
Similarly, they might discover one gluon exchange dominance for the
mass differences between baryons of the same quark constituency but
different spin.

\subsection{Beyond $\mathbf{q\bar q}$ and $\mathbf{qqq}$}

A currently very active endeavor is the analysis of hadrons with a larger quark
constituency than the consuetudinary old one\footnote{See e.g.~the talks by Marina Nielsen
and Sebastian Neubert at this conference \cite{Conf}.}. The theoretical prehistory of this subject
dates back to the mid 70's \cite{Molecules}. Perhaps the first data analyses with
{\it Molecular Charmonium} in mind were those of \cite{OurMolecules,DJ}.
In the first of these papers we concluded: {\it It seems very likely to 
us that four-quark molecules involving a $c\bar c$ pair do
exist, and have a rich spectroscopy. Our conjecture that the 4.028 GeV and
perhaps de 4.4 GeV peaks in $e^+e^-$ annihilation are indeed due to the
production of these molecules is more speculative. If it is true, then nature
has provided us with a spigot to a fascinating and otherwise almost inaccessible
new ``molecular'' spectroscopy full of experimental and theoretical challenges.}

\begin{figure}[h]
\centering
\sidecaption
\includegraphics[width=10.5cm]{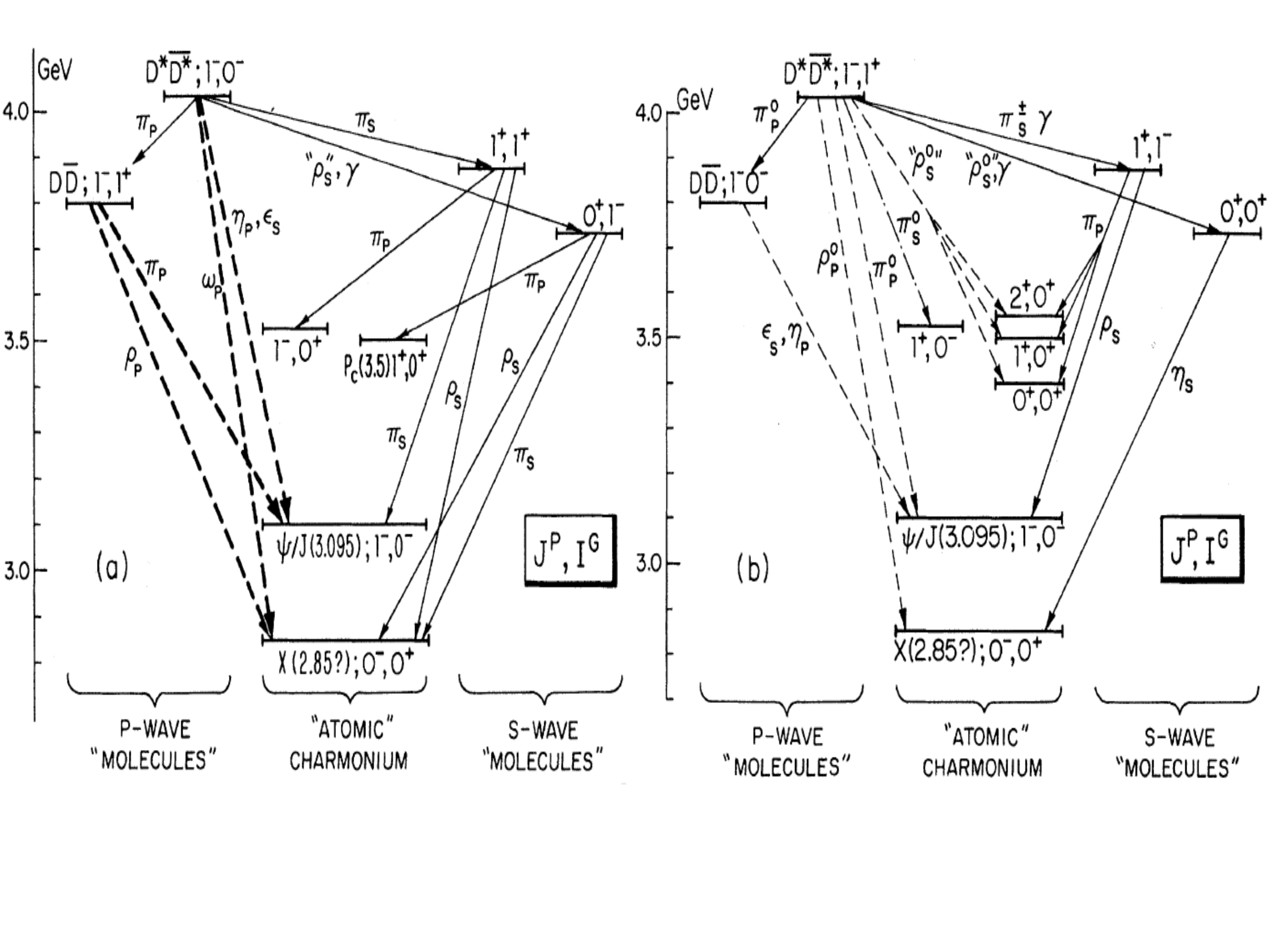}
\vspace{-1.5cm}
\caption{The spectrum of molecular charmonium in $e^+e^-$ annihilation above 3.6 GeV,
according to \cite{OurMolecules}.}
\label{fig-MolMess}   
\end{figure}

This conclusion is still unaltered, but for two details: 
our lack of prescience in the {\it almost inaccessible} stipulation, and the
possibility that some of the new hadrons are more ``nuclear'' than ``molecular''.
That the situation would be very messy, even in a narrow energy domain 
in $e^+e^-$ annihilation, could already be concluded from figure~\ref{fig-MolMess}.
I have no idea whether any of its predictions are correct.

\section{Back to the past}

In the summer of '75 --after a year of upper limits incompatible with
the theoretical expectations--
evidence finally arose for the $P$-wave
 charmonia  \cite{Pc}. The DESY experimentalists did
not refer to  the theorists who suggested their search; they are hereby
punished: they do not get a  reference, and they will remain
eternally ignorant of my juicy version of the story of their
competition with SLAC.

The discovery of the positronium-like $c\bar c$ spectrum of figure~\ref{fig-Charmonia} 
started to convert many infidels to the quarker faith. And the charmed quark, not
yet found unaccompanied by its antiparticle, was
to continue playing a crucial role in the development and general
acceptance of the standard lore.

\section{Yet Another Year of Lank Cows}

\begin{figure}[h]
\centering
\sidecaption
\includegraphics[width=10cm]{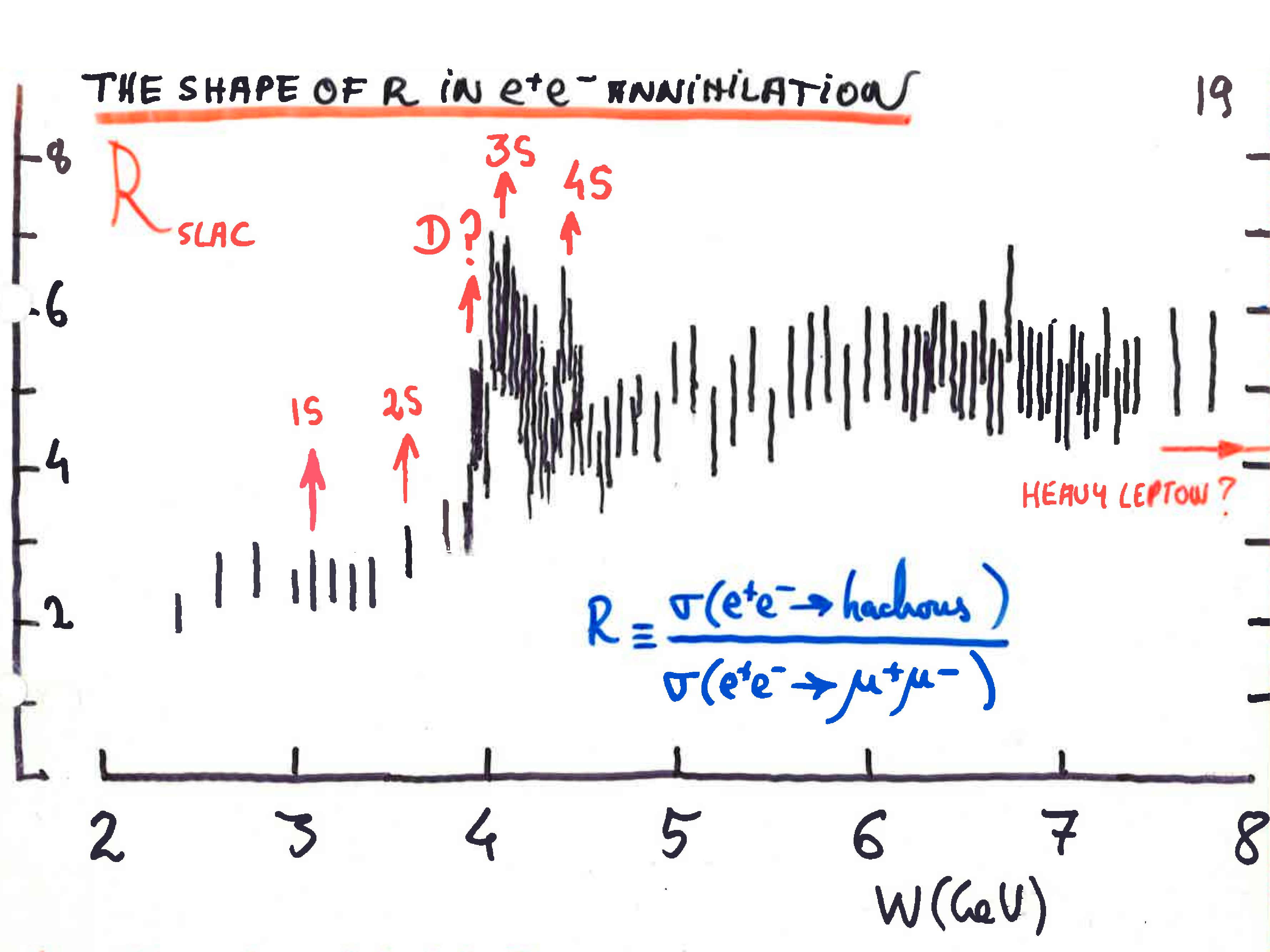}
\caption{A vintage transparency depicting $R$ in the interesting domain.}
\label{fig-RSLAC}       
\end{figure}

As shown in figure \ref{fig-RSLAC}, measurements at SLAC of the ratio 
$R\equiv \sigma(e^+e^-\to {\rm hadrons})/\sigma (e^+e^-\to  \mu^+\mu^-)$  
showed a doubling of the yield and structure aplenty as the $\sqrt{s}\sim 4$
GeV region is crossed \cite{Sig}.
Much of the jump {\bf had} to be due to the
production of charmed pairs, which were not found. Howard
Georgi and I innocently
believed that a serious sharpening of the arguments would help.

In the space-like domain, $s\!<\!0$, QCD predictions 
for $e^+e^-$ annihilation are insensitive
to thresholds, bound-state singularities and hadronization caveats.
For years, theorists had been unjustifiably applying the predictions
to the time-like domain wherein experimentalists insist on taking
$e^+e^-$ data. In a paper  \cite{HYo} whose rhythmic title {\it  
Finding Fancy Flavours Counting Colored Quarks} was duly censored, we
transferred the $e^+e^-$ data, via a dispersion relation, to a
theoretically safer space-like haven.  This somersault  \cite{Adler}
allowed us to conclude that {\it the old theory with no
charm is excluded, the standard model with charm is acceptable if
heavy leptons are produced, and six quark models are viable if no
heavy leptons are produced}. Thus, anybody listening to the other
voice in the desert (that of Martin Perl, who was busy demonstrating
that he had discovered the $\tau$) had no choice but 
charm. 

Our work was improved by Enrico Poggio, Helen Quinn and Steven Weinberg  \cite{PQW}, who
realized that one could, in the complex $s$-plane, work in a contour
around the real axis where perturbative QCD can still be trusted,
whilst the distance from the dirty details of real life is judged
safe. The work of Enrico, Helen and Steve further strengthened our conclusion: the
measured total cross section, analyzed on firm theoretical grounds,
implied the existence of charm and of a new heavy lepton.

Imagine that some theorists, analyzing LHC data with the current Standard
Model --with its six quarks and three charged leptons-- and on the basis of a ship-shape analysis
with a statistical evidence so strong that there was no need to count $\sigma$'s,
proved that an extra quark and an extra charged lepton were being produced.
There is no doubt that the community would conclude that the cited theorists
had discovered these particles. But the social power of preconceptions cannot be overestimated.
Prior to 1976, the Standard Model --then having three established quarks and two observed
charged leptons-- was not yet accepted as ``part of the truth''. Thus, to be believed,
the analyses in  \cite{HYo} and \cite{PQW},  had to wait for the explicit discovery of 
open charm and the $\tau$ lepton.

\section{Charm is found}

No amount of published information can
compete with a few minutes of conversation. The story, whose moral
that was, is well known. For the record, I should tell it once again
 \cite{Aachen,Goldhaber,RefA}:

Shelly Glashow happened to chat with  Gerson Goldhaber in an airplane.
Surprisingly, the East Coast theorist
managed to convince the West Coast experimentalist of something.
There was no way to understand the data unless charmed particles were
being copiously produced above $\sqrt{s}=3.7$ GeV. The
experimentalists devised an improved (probabilistic) way to tell
kaons from pions. In a record 18 days two complementary SLAC/LBL
subgroups found convincing evidence for a new particle  
with all the earmarks of charm \cite{Gerson}. The charmonium
advocates at Cornell had been trying for a long time to convince the
experimentalists to attempt to discover charm by sitting on the
$\psi(3440)$ resonance, or on what would become a ``charm factory":
$\psi(3770)$ \cite{Corncharm}. Alas, they initially failed.

\begin{figure}[h]
\centering
\sidecaption
\includegraphics[width=8cm]{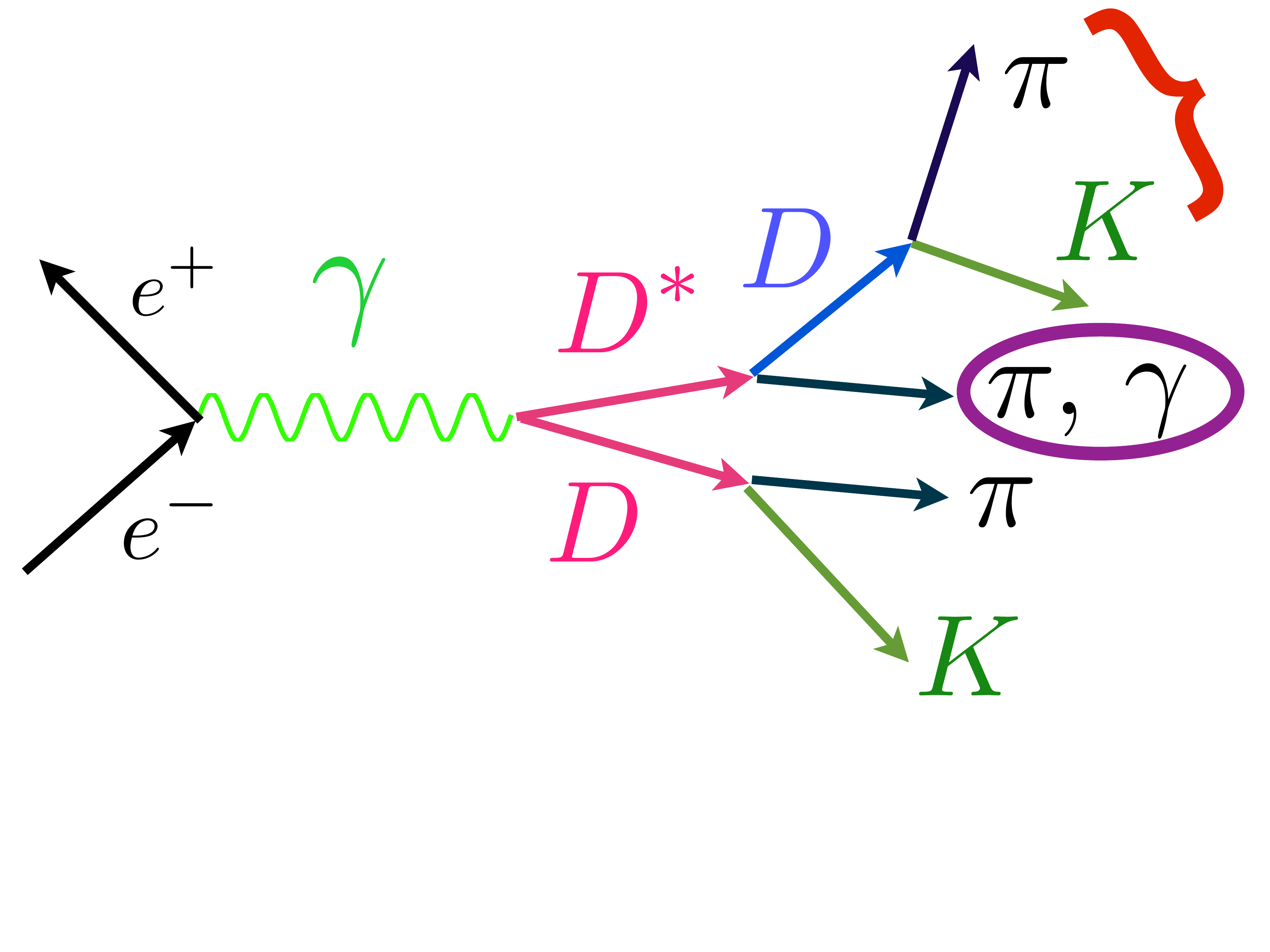}
\vspace{-1.5cm}
\caption{$\bar D\, D^*$ pair-production. Invariant masses (IM)
are measured as recoiling from $K\pi$ (as in the figure), or $K\pi\pi$ ensembles.
The decay $D^*\to D\pi$ is allowed, forbidden or suppressed, depending on the particle's
charges \cite{DGG}. Close to the open-charm threshold all observed
hadrons are ``slow" and fake IM peaks consequently occur.}
\label{fig-CP}       
\end{figure}

The observation of charmed mesons
ought to have been the immediate happy ending, but there was a last-minute
delay. The invariant-mass spectrum of recoiling stuff in $e^+e^-\to
D^0\,(D^\pm)+...$ had a lot of intriguing structure, but no clear
peak corresponding to $D^0\bar D^0$  \cite{Gerson} or $D^+D^-$
 \cite{Peru} associated production. Enemies of the people rushed to
the conclusion that what was being found was an awful mess, and not
something as simple as charm, as in figure~\ref{fig-CP} .

\begin{figure}[h]
\centering
\sidecaption
\includegraphics[width=10.5cm]{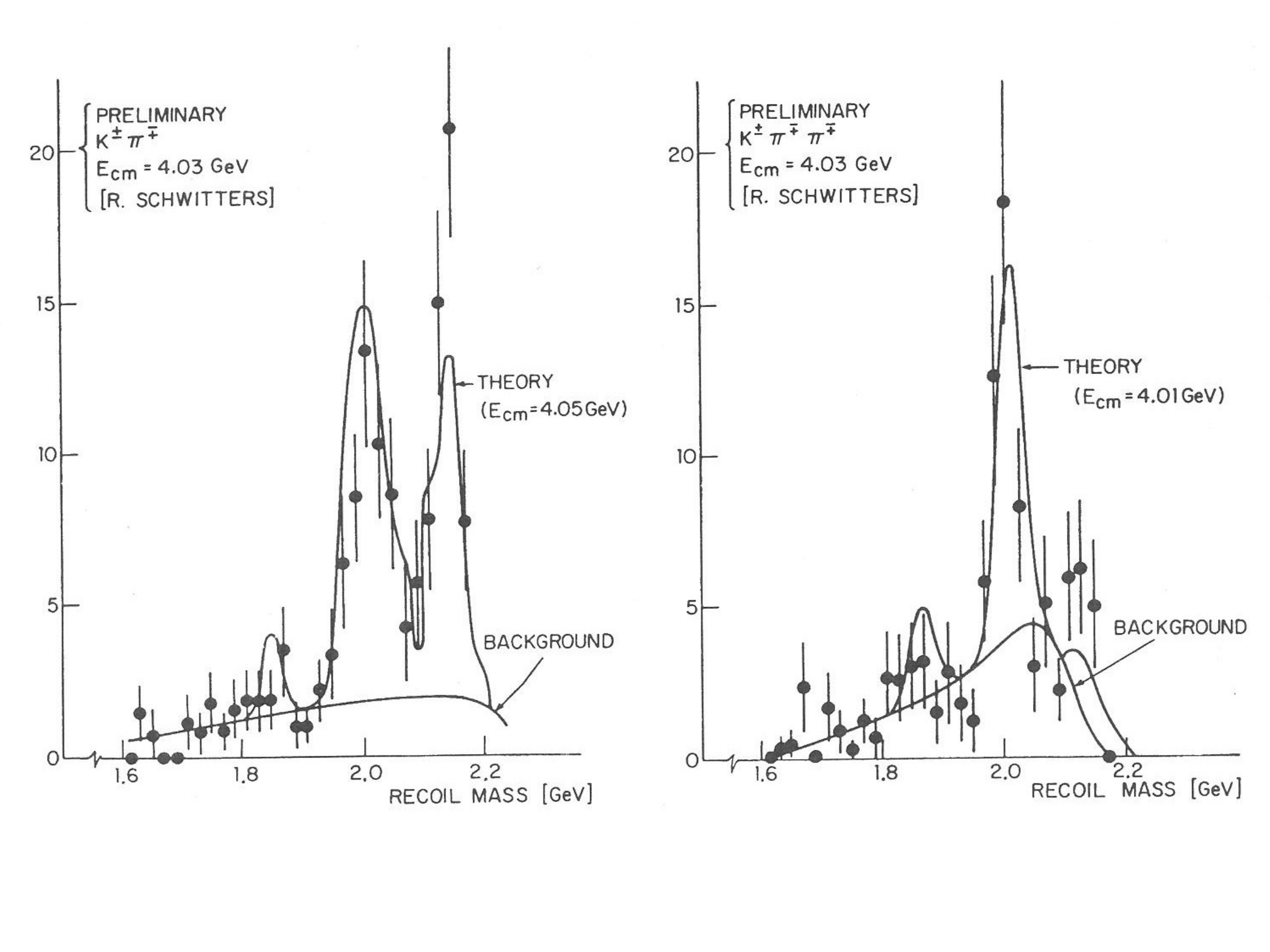}
\vspace{-1.5cm}
\caption{Predicted and observed invariant-mass spectra, recoiling
against neutral and charged $D$'s. The theoretical
curves are a one-parameter description  \cite{Charm}.}
\label{fig-CD}       
\end{figure}

But we had one  last unspent cartridge  \cite{Charm}. We expected $D\bar
D$, $D\bar D^*+\bar D D^*$, and $D^*\bar D^*$ production to occur in
the ``spin'' ratio 1:4:7 (thus the $D\bar D$ suppression). We trusted
our prediction  \cite{DGG} $m(D^*)-m(D)\simeq m(\pi)$, which implies
that for charm production close to threshold, the decay pions are
slow and may be associated with the ``wrong'' $D$ or $D^*$ to produce
fake peaks in recoiling mass. Finally, we knew that the charged $D$'s
and $D^*$'s ought to be a little heavier than their neutral sisters.
The $D^*$ decays had to be very peculiar: $D^{*0}\to D^+\pi^-$ is
forbidden, $D^{*0}\to D^0\gamma$ competes with $D^{*0}\to D^0\pi^0$,
etc. On the basis of these considerations (and with only one fit
parameter) we constructed the recoil spectra shown in figure~\ref{fig-CD}.
 Case closed!

\section{Conclusion}

In the olden days experimentalists, particularly the ones working in the 
East Coast of the USA, were strongly motivated to disprove all
theories and to mistrust almost all theorists; they were
perhaps permeated by some arcane Californian
faith that nature is intrinsically unfathomable. They definitely
did not have in their data-analysis programs the current instruction
stating: IF [RESULT = STANDARD; LOOK ELSEWHERE]. This made life
most enjoyable and the case for the then-challenged  standard model {\it veeeery} strong.
To gauge whether or not things have changed, consider supersymmetry.

Even the most formal theorist or the most cable-connecting experimentalist
understands positronium and hydrogen. These objects are not so very different
from their QCD analogs, charmonium
and charmed particles. This may be why it took asymptotic freedom and a
fourth (charmed) quark to have the Standard Model become the standard lore.

\section{Windup}

The most challenging QCD problem --confinement-- is not yet solved,
in spite of a one M\$ prize awaiting  whoever solves it \cite{Clay}.
Factorization in deep inelastic scattering (the fact that one can use parton distribution
functions) is proven \cite{Fac}. But it is not proved for Drell-Yan processes, a crucial basis in
the analyses of LHC data. Nowadays theorists are also considering
processes allegedly initialized by multiple partons. This is even less solidly founded.
And there is no prize for clarifying any of this. 

Independently of the above caveats, as this conference reflected quite well,
there has been an enormous progress in perturbative and non-perturbative QCD
since quarks were invented, in 1963.
This phenomenological progress often required phenomenally difficult
theoretical developments, and played a key role in the understanding of experimental
results, most recently at the LHC. It is difficult not to feel that particle physics
phenomenology is generally less appreciated than it should,  particularly
in comparison with theoretical work ``beyond'' this or that. A reader who has
reached this far is presumably a QCD phenomenologist. I have no doubt
that (s)he would agree with me on all this.

\vspace{.6cm}

\noindent {\bf Acknowledgments.} 
I am most indebted to my collaborators, in particular the ones I have
cited. In alphabetical order: Tom, Howard, Shelly and David.
I am also thankful  to Johannes Bluemlein, Cesareo Dominguez, Mario Greco,
Marek Karliner and Martinus Veltman for comments and suggestions.
This project has received funding from the European Union's Horizon 2020 
research and innovation programme under the Marie Sklodowska-Curie 
grant agreement No 674896.
%

%

%For bibliography use \cite{RefJ}

%Don't forget to give each section, subsection, subsubsection, and
%paragraph a unique label (see Sect.~\ref{sec-1}).

%
%For figure with sidecaption legend use syntax of figure~\ref{fig-2}.

%
%For tables use syntax in table~\ref{tab-1}.
%\begin{table}[h]
%\centering
%\label{tab-1}       % Give a unique label
%% For LaTeX tables you can use
%\begin{tabular}{lll}
%\hline
%first & second & third  \\\hline
%number & number & number \\\hline
%\end{tabular}
%% Or use
%%\vspace*{5cm}  % with the correct table height
%\end{table}
%%
%% BibTeX or Biber users please use (the style is already called in the class, ensure that the "woc.bst" style is in your local directory)
%% \bibliography{name or your bibliography database}
%%
%% Non-BibTeX users please use
%%

%%%%%%%%%%%The figures and tables must be before references.

%\clearpage

\end{document}